\documentclass[12pt,preprint]{aastex}

\slugcomment{To appear in ApJ}

\shorttitle{DIBs in the RCB DY Cen}
\shortauthors{Garc\'{\i}a-Hern\'andez et al.}

\begin{document}


\title{High-resolution optical spectroscopy of DY Cen: diffuse interstellar
bands in a proto-fullerene circumstellar environment?}


\author{D. A. Garc\'{\i}a-Hern\'andez\altaffilmark{1,2}, N. Kameswara
Rao\altaffilmark{3,4} and David L. Lambert\altaffilmark{4}}


\altaffiltext{1}{Instituto de Astrof\'{\i}sica de Canarias, C/ Via L\'actea
s/n, 38200 La Laguna, Spain; agarcia@iac.es}
\altaffiltext{2}{Departamento de Astrof\'{\i}sica, Universidad de La Laguna (ULL), E-38205 La Laguna, Spain}
\altaffiltext{3}{543, 17$^{th}$ Main, IV Sector, HSR Layout, Bangalore 560102, (Indian Institute of Astrophysics, Bangalore 560034,) India; nkrao@iiap.res.in}
\altaffiltext{4}{The W.J. McDonald Observatory, University of Texas, Austin, TX 78712-1083, USA; dll@astro.as.utexas.edu}


\begin{abstract}
We search high-resolution and high-quality  VLT/UVES optical spectra of the hot
R Coronae Borealis (RCB) star DY Cen  for  electronic transitions of the
C$_{60}$  molecule and diffuse interstellar bands (DIBs). We report the
non-detection of the strongest C$_{60}$ electronic transitions (e.g., those at
$\sim$3760, 3980, and 4024 \AA). Absence of C$_{60}$ absorption bands may
support recent laboratory results, which show that the $\sim$7.0, 8.5, 17.4, and
18.8 $\mu$m emission features seen in DY Cen - and other similar objects with
PAH-like dominated IR spectra - are  attributable to proto-fullerenes or
fullerene precursors rather than to C$_{60}$. DIBs towards DY Cen are normal for
its reddening; the only exception is the DIB at 6284 \AA\ (possibly also the
7223\AA\ DIB) that is found to be unusually strong. We also report the detection
of a new broad (FWHM$\sim$2 \AA) and unidentified feature centered at $\sim$4000
\AA. We suggest that this new band may be related to the circumstellar
proto-fullerenes seen at infrared wavelengths. 
\end{abstract}


\keywords{astrochemistry --- circumstellar matter --- ISM: molecules --- dust,
extinction --- Stars: chemically peculiar --- Stars: individual: (DY Cen)}



\section{Introduction}

Fullerenes and fullerene-related molecules have attracted much attention since
their discovery in the laboratory (Kroto et al. 1985) due to their potential
applications in superconducting materials, optical devices, and medicine. In the
astrophysical context, these complex molecules have been proposed as
explanations for unidentified astronomical features such as the intense UV
absorption band at 217 nm (e.g., Cataldo \& Iglesias-Groth 2009) and the
enigmatic diffuse interstellar bands (DIBs) (Herbig 1995, 2000; Iglesias-Groth
2007). The remarkable stability of fullerenes against intense radiation (e.g.,
Cataldo, Strazzulla \& Iglesias-Groth 2009) has encouraged the idea that
fullerenes should be present in the interstellar medium. Indeed, the 9577 and
9632 \AA\ DIBs observed in a few reddened stars lie near two electronic
transitions of the C$_{60}$ cation observed in rare gas matrices (Foing \&
Ehrenfreund 1994) but the presence of fullerenes in astrophysical environments
has been uncertain until  recently, when the mid-IR spectral signatures of the
C$_{60}$ and C$_{70}$ fullerenes were  detected in several Planetary Nebulae
(Cami et al. 2010; Garc\'{\i}a-Hern\'andez et al. 2010, 2011c). 

Interestingly, Garc\'{\i}a-Hern\'andez et al. (2010) showed  that contrary to
theoretical and experimental expectation, fullerenes  are efficiently formed in
normal H-rich circumstellar environments. The prevailing laboratory view was
that the synthesis of fullerenes is much more efficient under hydrogen-poor
conditions (Kroto et al. 1985; De Vries et al. 1993). Thus, circumstellar
envelopes around R Coronae Borealis (RCB) stars (see e.g., Lambert \& Rao 1994
for a review) have been considered as promising environments for the formation
of fullerene molecules (Goeres \& Sedlmayr 1992). This is because the  RCBs'
hydrogen deficiency together with the He- and C-rich character of the gas
resembles the laboratory conditions  where fullerenes are produced. However,
this suggestion is not supported by our recent {\it Spitzer} observations of a
large sample of RCB stars, which show that fullerene formation is inefficient in
the highly H-deficient environments characteristic of most RCBs
(Garc\'{\i}a-Hern\'andez et al. 2011a,b). 

Surprisingly, the only exception among RCBs is the least H-deficient RCB star DY
Cen (possibly also the second least H-deficient RCB star V854 Cen) that shows
mid-IR emission features at $\sim$7.0, 8.5, 17.4, and 18.8 $\mu$m coincident
with the IR transitions of C$_{60}$. Contrary to fullerene-containing PNe, the
latter C$_{60}$ mid-IR features in DY Cen are seen in conjunction with very
strong polycyclic aromatic hydrocarbon (PAH) features. This striking difference
between the IR spectra of fullerene-containing PNe and DY Cen is explained by
Duley \& Hu (2012) who attribute the $\sim$7.0, 8.5, 17.4, and 18.8 $\mu$m
features in DY Cen to proto-fullerenes (or fullerene precursors)  rather than to
the C$_{60}$ molecule. Based on their laboratory spectroscopy of hydrogenated
amorphous carbon (HAC) nanoparticles, Duley \& Hu (2012) suggest that these four
IR features may be due to proto-fullerenes in sources such as reflection
nebulae, RCB stars, and proto-PNe, which also show the unidentified 16.4 $\mu$m
feature and other very strong PAH-like features. In light of the now suspect
infrared identification of neutral C$_{60}$, we sought to detect the molecule's
electronic transitions in absorption from DY Cen's optical spectrum, as an
additional  check to the
presence of C$_{60}$ and see whether  Duley \& Hu (2012)'s ``proto-fullerenes" 
suggestion could be substantiated. 

\section{VLT/UVES optical spectroscopy}

High-resolution (R$\sim$30,000) and high-quality (signal-to-noise ratio
S/N$\geq$200) optical ($\sim$3300-9450 \AA) spectroscopic observations of DY Cen
(V=12.7, E(B-V)=0.50, see Table 1) were obtained in the period February-March
2010 at the European Southern Observatory (ESO).  The UVES spectrograph at the
ESO VLT (Paranal, Chile) was used with the 1.2" slit and the standard setting
DIC2 (390+760). We obtained 11 individual exposures of 1800 s each, giving a
total exposure time of 5.5 hours. The S/N in the continuum in the summed DY
Cen's spectrum is $\sim$200 at 4000 \AA\ and higher than 250 at wavelengths
longer than 6000 \AA. The observed spectra - processed by the UVES data
reduction pipeline (Ballester et al. 2000) - were corrected for heliocentric
motion and the stellar continuum was fitted by using standard astronomical tasks
in IRAF. 

As a comparison star for DY Cen in our search for C$_{60}$ and DIBs, we selected
the very nearby B0.5 Ia bright supergiant HD 115842, which only differs in
Galactic longitude and latitude by 0$^o$.88 and 1$^o$.46, respectively. This is
because HD 115842 has the same reddening (E(B-V)=0.5) as DY Cen and it may be
located at a comparable distance, sampling similar ISM conditions. A comparison
between HD 115842 and DY Cen will probably enlighten the environmental changes
associated with DY Cen. Thus, HD 115842 was also observed on the same dates as
DY Cen with the same VLT/UVES set-up. Table 1 compares DY Cen with the
comparison star HD 115842 in terms of Galactic coordinates, spectral type,
magnitude, and reddening. Finally, a spectrum of BD $-$9$^\circ$ 4395, an
extreme helium (EHe) star that shows a photospheric spectrum similar to that of
DY Cen, was obtained by us with the 2.7m Harlan Smith telescope and the Tull 
coud\'{e} spectrograph (Tull et al. 1995)  at a spectral resolving power of
60,000. 

\section{A search for neutral C$_{60}$}

We have inspected the optical spectrum of DY Cen in order to search for the
presence of optical absorptions of neutral C$_{60}$. The allowed transitions at
wavelengths below 4100 \AA\ are much more intense than those of the
Herzberg-Teller induced forbidden transitions at longer wavelengths (Leach 1992).
The strongest  allowed electronic transitions of neutral gas phase C$_{60}$
molecules, as predicted from laboratory experiments, are located at  $3760\pm5$,
3980$\pm0.5$, and 4024.0$\pm0.5$ \AA\ with widths of 8, 6, and 4 \AA,
respectively (Sassara et al. 2001). The oscillator strength (f) of the $\sim$4024
C$_{60}$ band is similar to that at 3980 \AA\ (Leach 1992),  while the
3760 \AA\ band has a five times higher oscillator strength (Braga et al. 1991). 
DY Cen's optical spectrum is dominated by strong  He I, C II, Ne I, O II, and N
II absorption and variable emission lines  together with nebular forbidden
emission lines from [O II], [N II], [S II], etc. The presence of these absorption
and emission lines (FWHM$\sim$1 \AA) complicates the detection of broad and weak
absorption features. 

We can find no evidence for the presence of C$_{60}$ in  absorption (or
emission) at the wavelengths of the expected  electronic transitions: \\ {\bf
The 3760${\bf\pm5}$ \AA\ band:} This, the strongest of the three in our
bandpass,  falls amongst a series of O II lines. These lines give consistent O
abundances suggesting that none is does blended with a C$_{60}$ feature (Figure
1, left panel), even though its central wavelength is uncertain by $\pm5$ \AA\
(Sassara et al. 2001). Note that the apparent additional absorption at 3756
\AA\ in DY Cen is a stellar blend (Figure 1, left panel).

\noindent {\bf The 3980${\bf\pm0.5}$ \AA\ band:} This  coincides with a stellar
C II line from a multiplet represented by other lines. A secondary contributor
may be the Al III 5d $^2$D - 8f $^2$F$^\circ$ multiplet (Figure 1, right
panel). The  C II lines   are present also in BD $+9^\circ 4395$. This
coincidence and the line's `stellar' width show that C$_{60}$ is not a
significant contributor.  An interesting feature of the right panel of
Figure 1 is the appearance of an unidentified broad feature in DY Cen at 4000
\AA\ (FWHM $\sim$ 2 \AA; equivalent width of 91 m\AA). It is to be noted
here that the 4000 \AA\ feature is real. First, a strong argument in favour of
the 4000 \AA\ feature being real is that it doest not appear in the spectra of
the comparison star taken with the same set-up and at the same time. Second, the
4000 \AA\ band is clearly seen when reducing the corresponding {\it Echelle}
order alone. This is not a known DIB; it's not present in HD 115842 where all
DIBs have a similar strength to those in DY Cen (see below) and neither is it
listed by Hobbs et al. (2008). Identification as C$_{60}$ would demand a
wavelength error of either 20 \AA\ for the 3980 \AA\ transition or 24 \AA\ for
the 4024 \AA\ transition but Sassara et al. give the wavelength uncertainty as
just $\pm0.5$ \AA\ for both transitions.

\noindent {\bf The C$_{60}$ band at 4024.0${\bf\pm0.5}$:} This is irretrievably
blended with the He I 4026 \AA\ line and adjacent lines. 

We estimate that the one-sigma detection limits on the column density from
the 3760 and 3980 \AA\ C$_{60}$ bands in our DY Cen spectra (S/N$\sim$200) are
4.6 $\times$ 10$^{12}$ and 1.5 $\times$ 10$^{13}$ cm$^{-2}$,
respectively\footnote{1-sigma detection limits for the EQWs in our spectra
scale as $\sim$ 1.064  x FWHM / (S/N) (see e.g., Hobbs et al. 2008).}. It is of
interest to compare this column density  limit of about 10$^{13}$ cm$^{-2}$ with
the number of C$_{60}$ molecules calculated from the infrared emission features
attributed to these molecules. This exercise is attempted next.

The IR features measure the number of molecules in excited vibrational states
but the electronic transitions sought here depend on the molecules in their
ground vibrational state. Analysis of the strengths of the IR features with the
estimated A-values (Cami et al. 2010; Garc\'{\i}a-Hern\'{a}ndez 2011a) shows
that the excitation temperature is about 600 K. With this temperature, we
estimate the number of molecules in the ground state: $N(C60) = 1.0 \times
10^{46} d^2$ where $d$, the distance to DY Cen, is given in kpc.

The column density pertinent to our non-detection of C$_{60}$ depends on how the
line of sight to the star intersects  the C$_{60}$-containing part of the
circumstellar shell. Assume a spherical shell centered on the star with inner and
outer radii of $R_{out}$ and $R_{in}$, respectively, with the line of sight
running directly from $R_{out}$ to $R_{in}$.  Assume a uniform density of
C$_{60}$ molecules throughout the shell, then the density of molecules is
$n$(C$_{60}$) where

\begin{equation}
 n(C_{60}) = \frac{7.0 \times 10^{-41}}{R^3_{out} - R^3_{in}} N(C_{60})
\end{equation}

\noindent where $R_{out}$ and $R_{in}$ are given in AU.

Shell radii are estimated from the run of dust temperature with distance
from the star. The equilibrium temperature of a gray dust grain $T_d(R)$ in
an optically thin circumstellar environment is given by

\begin{equation}
T_d(R) = \left(\frac{R_*}{2R}\right)^{0.5} T_*
\end{equation}

\noindent where $R_*$ is the stellar radius, R is the radial distance from the
stellar center, and $T_*$ is the stellar blackbody temperature
(Kwok 2007, p. 314, Equation (10.32)). If stellar radii and temperature are
estimated from the stellar luminosity $L/L_{\odot}$, the distance $R$ is provided
by

\begin{equation}
R = \frac{1}{430} \left(\frac{L}{L_\odot}\right)^{0.5} \left(\frac{T_\odot}{T_d(R)}\right)^2
\end{equation}

\noindent for $R$ in AU.

Adopting $L/L_\odot = 10^4$, we find $T_d$ = 600 K at $R$= 21 AU and
$T_d$ = 270 K at 104 AU where the former temperature is the excitation
temperature of the C$_{60}$ molecules and the latter temperature is
the blackbody temperature of the dust. If we adopt R$_{in}$ $\simeq 10$ AU
and $R_{out} \simeq 100$ AU, we obtain $n$(C$_{60}$) = 0.7d$^{2}$ cm$^{-3}$
or $n$(C$_{60}$) = 34 cm$^{-3}$ for $d = 7$ kpc and 17 cm$^{-3}$ for $d = 5$
kpc. Then, the estimates for the C$_{60}$ column density along the
path ($R_{out} - R_{in}$) are $4.6 \times 10^{16}$ cm$^{-2}$ for 7 kpc and
$2.3 \times 10^{16}$ cm$^{-2}$ for 5 kpc.

Such estimates are uncomfortably greater than the observed upper limit of
around $10^{13}$ cm$^{-2}$. One possible explanation of this 1000-fold
discrepancy is that the line of sight to DY Cen does not intersect the
C$_{60}$-containing regions of the circumstellar envelope. An alternative
view  raises  doubts about the IR detection of the C$_{60}$
molecule in DY Cen (possibly also in V854 Cen; Garc\'{\i}a-Hern\'andez et al.
2011a), supporting the recent claim that the $\sim$7.0, 8.5, 17.4, and 18.8
$\mu$m features seen in astronomical sources with PAH-like dominated IR spectra
should be attributed to proto-fullerenes or fullerene precursors rather than to
C$_{60}$ (Duley \& Hu 2012). The latter seems to be consistent with the
characteristics of the DIBs towards DY Cen, which are studied below.

\section{Diffuse bands from fullerene precursors?}

As we have mentioned above, we detect in DY Cen a broad unidentified feature
centered at $\sim$4000 \AA, which is seen in DY Cen only (Fig. 1). Note that no
diffuse interstellar bands (DIBs) are known at this wavelength (see e.g., Hobbs
et al. 2008). To our knowledge, no molecule is known to exhibit a strong
electronic transition at $\sim$4000 \AA\ and this is corroborated by inspecting
the NIST Chemistry WebBook Database\footnote{See
http://webbook.nist.gov/chemistry/.}. We speculate that this 4000 \AA\ band may
be related to the proto-fullerenes or fullerene precursors seen in the IR and
that could be very abundant in the circumstellar envelope of DY Cen. Thus, the
4000 \AA\ feature may represent a new and very unusual diffuse band (DIB). 

We used the exhaustive list of DIBs provided by Hobbs et al. (2008) in HD 204827
to search for them in the spectrum of  DY Cen.  DIBs detected in DY Cen (and the
comparison star HD 115842) are given in Table 2, where we give the central
wavelength (after applying the heliocentric correction), full width at
half-maximum (FWHM) in \AA, equivalent width (EQW) in m\AA, central depth
(A$_{c}$), and the S/N ratio in the neighbouring continuum. Figure 2 compares
a selection of DIBs in DY Cen with those in HD 115842.  Table 2 (also
Figure 2) shows that the DIBs towards DY Cen have about the same
or slightly less (up to 30 per cent)  strength, the same FWHM, and a similar
radial velocity as those towards HD 115842. Since  both stars have roughly the
same interstellar reddening E(B-V), the similar profiles and strengths for their
DIBs is not surprising.  In addition, for the  well studied DIBs at 5780, 5797,
5850, 6196, 6379, and 6614 \AA, the normalised equivalent widths EQW/E(B-V) in
DY Cen agree very well with those measured in field stars by Luna et al. (2008)
and in the prototype star HD 183143  (Herbig 1995). 

The exceptions to the above trends are the DIBs at 6284 \AA\ and at 7223 \AA.
The 7223\AA\ DIB is stronger in DY Cen than in HD 115824 as seen from the
divided spectrum, but we could not estimate its total absorption because of the
interference from the superposed telluric lines. Figure 3 (left panel) shows the
region of 6284 \AA\ for both stars. The spectra have been cleaned of telluric
lines. It is clear that this DIB towards DY Cen is stronger than towards HD
115842, suggesting that the carrier of the 6284\AA\ DIB (along with 7223 \AA) is
different from the rest of the classical DIBs.  Indeed, the strength of this DIB
seems to be poorly correlated with the interstellar reddening. For example, Luna
et al. (2008) and Herbig (1995) obtained EQW/E(B-V) of 0.90 and 1.50 for field
reddened stars and HD 183143, respectively, while we measure a value of 1.10
towards DY Cen; Hobbs et al. (2008) give a value of 0.41 in HD 204827.
Curiously, the post-AGB star IRAS 06530$-$0213, which shows the strongest
unidentified 21 $\mu$m feature (Zhang, Kwok \& Hrivnak 2010), also shows
an unusually strong 6284\AA\ DIB (see Luna et al. 2008). The 21 $\mu$m feature
may be attributed to HACs or their decomposition products such as fullerene
precursors or intermediate products (see e.g., Garc\'{\i}a-Hern\'andez 2012). It
is to be noted here that examination of atomic and  molecular lines towards the
sight lines of DY Cen and HD 115842 show that ionized lines are enhanced in
strength in the sight line towards DY Cen relative to that of HD 115824, which
might suggest that the carrier(s) of the 6284\AA\ and 7223\AA\ DIBs could even
have some contribution from ionized species. 

Finally, we would like to remark that the very broad 4430 \AA\ DIB in DY Cen is
almost identical to that in the comparison star HD 115842. This is shown in the
righthand panel of Figure 3 where we compare the 4400$-$4450 \AA\ spectral
region for both stars. The 4430 \AA\ DIB has been linked to fullerenes bigger
than C$_{60}$ (e.g., C$_{80}$, C$_{240}$, C$_{320}$, and C$_{540}$) and/or
buckyonions such as C$_{60}$@C$_{240}$ and C$_{60}$@C$_{240}$@C$_{540}$
(Iglesias-Groth 2007). Our finding also would be consistent with fullerenes and
fullerenes-containing molecules not being especially overabundant towards DY
Cen.

\section{Concluding Remarks}

Duley \& Hu (2012) show that the infrared features at $\sim$7.0, 8.5, 17.4, and
18.8 $\mu$m detected in astronomical sources with PAH-like dominated IR spectra
- such as those of the RCB stars DY Cen and V854 Cen (Garc\'{\i}a-Hern\'andez et
al. 2011a), the Reflection Nebulae NGC 7023 and NGC 2023 (Sellgren et al. 2010)
or the proto-PN IRAS 01005$+$7910 (Zhang \& Kwok 2011) should be attributed to
fullerene precursors (or proto-fullerenes) rather than to C$_{60}$. Our non
detection of neutral C$_{60}$ in DY Cen may support this claim . Thus, the
carrier of the new 4000 \AA\ absorption band detected in DY Cen may be
intimately related with the fullerene precursors seen at infrared wavelengths.
Duley \& Hu (2012) suggest that fullerene precursors contain pentagonal carbon
rings; the infrared emission features at $\sim$7.0 and 16.4 $\mu$m usually
detected in objects with PAH-like dominated IR spectra are characteristics of
pentagonal rings (Moutou et al. 2000). Pentagonal carbon rings are present in
carbon nanoparticles (e.g., HACs; Duley \& Hu 2012) and nanotubes although these
have less perfect structures than C$_{60}$ and other fullerenes.

In summary, the non detection of neutral C$_{60}$ in the high-quality VLT/UVES
DY Cen's spectrum may support recent experimental work, showing that the
$\sim$7.0, 8.5, 17.4, and 18.8 $\mu$m IR features seen in sources with PAH-like
dominated spectra have to attributed to proto-fullerenes rather than to neutral
C$_{60}$. In addition, the new 4000 \AA\ DIB reported here (possibly also the
carriers of the classical 6284\AA\ and 7223\AA\ DIBs) may be related to fullerene
precursors; an organic compound containing pentagonal rings. These pentagonal
carbon rings are usually present in HAC nanoparticles and nanotubes, suggesting
that they may be intimately related with the formation process of fullerenes.



\acknowledgments

This work is based on observations obtained with the ESO Very Large Telescope at
Paranal under the Director's Discretionary Time (DDT) programme 284.D- 5048(A)
(PI: D. A. Garc\'{\i}a-Hern\'andez).  D.A.G.H. acknowledges support  provided by
the Spanish Ministry of Economy and Competitiveness under grant
AYA$-$2011$-$27754. D.A.G.H. also thanks his son Mateo for his great patience
during the realization of this work. D.L.L. also wishes to thank the Robert A.
Welch Foundation of Houston, Texas for support through grant F-634.



{\it Facilities:} \facility{VLT:UVES}.

\clearpage

\begin{table}
\centering
\begin{minipage}{110mm}
\caption{\Large  Observational parameters of  DY Cen. }
\begin{tabular}{llrrllllll}
\hline

Star & l & b & V & B-V  & E(B-V) & Sp. Type & $M_{\rm V}$ & Distance & Ref \\
    &   &   &   &      &        &   &          & (pc)     & \\
\hline
DY Cen & 307.958 & 8.293 & 12.7 & 0.33 & 0.47-0.50 & B0 Ia &   & 4800 & 1 \\
HD115842 & 307.080 & 6.834 & 6.04 & 0.30 & 0.50    & B0.5 Ia & -8.0 & 3581 &  \\
         &         &       &      &      &         & &-6.9 & 2157 & 2 \\
\hline
\end{tabular}
 1:- De Marco et al(2002), 2: Krelowski et al (2010), Hunter et al (2006)
\label{default}
\end{minipage}
\end{table}

\clearpage

\begin{table}
\centering
\begin{minipage}{120mm}
\caption{Diffuse Interstellar Bands in DY Cen and HD 115842$^{a}$}
\small\begin{tabular}{llllllllll}
\hline\hline
   \underline{DY Cen }&  &  & & &  \underline{HD 115842}& & & &  \\

 $\lambda$$_{c}$  & FWHM & EQW & Ac & S/N & $\lambda$$_{c}$ & FWHM & EQW & Ac & S/N \\
  (\AA) & (\AA) &  (m\AA) &     &       &  (\AA)   & (\AA)     & (m\AA)  &     &          \\
\hline
  4000.38 &  1.92 &  91 & 0.04 & 188 &         &      &     &      & 412 \\
  4429.66 & 21.10 & 1034& 0.05     & 273 &  4428.50 & 23.00     & 932    & 0.04    & 436    \\
  5780.46 & 2.14 & 256 & 0.10 & 209 & 5780.39 & 2.10 & 257 & 0.11 & 307 \\
  5796.92 & 0.91 & 72  & 0.07 & 199 & 5796.90 & 0.89 & 105  & 0.11 & 264 \\
  5849.67 & 0.82 & 32  & 0.02 & 272 & 5849.58 & 0.83 & 43  & 0.05 & 330  \\
  6065.11 & 0.61 &  6  & 0.01 & 312 & 6065.00 & 0.64 &  7  & 0.01 & 635  \\
  6089.62 & 0.53 &  7  & 0.01 & 282 & 6089.59 & 0.53 & 13  & 0.02 & 490  \\
  6112.7  & 1.00 & 12  & 0.01 & 300 & 6112.97 & 0.62 & 12  & 0.02 & 490  \\
          &      &     &      &     & 6116.62 & 0.94 &  8  & 0.01 & 490  \\
  6139.75 & 0.82 &  5  & 0.01 & 267 & 6139.72 & 0.59 &  8  & 0.01 & 570  \\
  6195.80 & 0.60 & 27  & 0.04 & 284 & 6195.80 & 0.50 & 33  & 0.06 & 636 \\
  6202.81 & 1.28 & 62  & 0.04 & 336 & 6202.83 & 1.35 & 64  & 0.04 & 636 \\
  6233.79 & 0.79 & 10  & 0.01 & 336 & 6233.80 & 0.66 & 14  & 0.02 & 560 \\
  6269.70 & 1.21 & 36  & 0.02 & 276 & 6269.80 & 1.38 & 70  & 0.04 & 582 \\
  6283.53 & 4.29 & 543 & 0.12 & 276 & 6283.75 & 4.43 & 413  & 0.09 & 582 \\
  6375.97 & 0.92 & 13  & 0.02 & 273 & 6375.91 & 0.90 & 29  & 0.03 & 340  \\
  6379.25$^{b}$ & 1.04 & 53  &      &     & 6379.04 & 0.64 & 70  & 0.10 & 340  \\
  6597.26 & 0.53 &  8  & 0.01 & 295 & 6597.19 & 0.58 &  9  & 0.01 & 378  \\
  6613.41 & 1.03 & 83  & 0.08 & 295 & 6613.45 & 1.04 & 149 & 0.14 & 304 \\
  6660.60 & 0.64 & 12  & 0.02 & 314 & 6660.53 & 0.80 & 39  & 0.05 & 265  \\
  7223.75 &      &     &      &     &         &      &     &      &      \\
  8620.88 & 3.75 & 95  & 0.02 & 106 & 8620.94 & 3.92 & 105 & 0.02 & 243  \\
\hline\hline
\end{tabular}
$^{a}$ The 3-$\sigma$ erros in the EQWs scale like $\sim$3 x FWHM / (S/N) while
we estimate that the FHWMs are precise to the 0.03 \AA\ level; 
$^{b}$\ blended with N II 6379.98 \AA\
\end{minipage}
\end{table}

\clearpage

\begin{figure}
\includegraphics[angle=0,scale=.45]{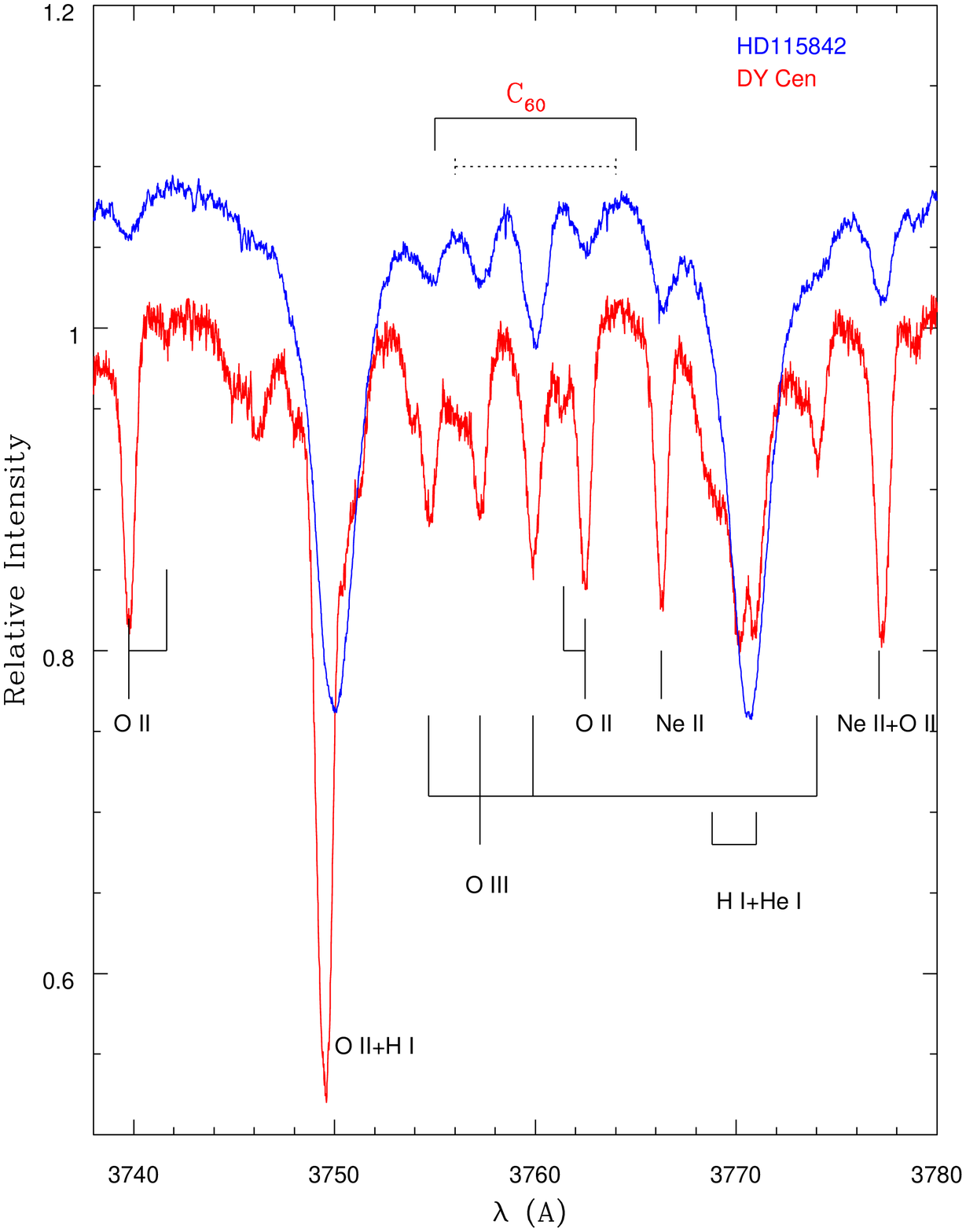}%
\includegraphics[angle=0,scale=.45]{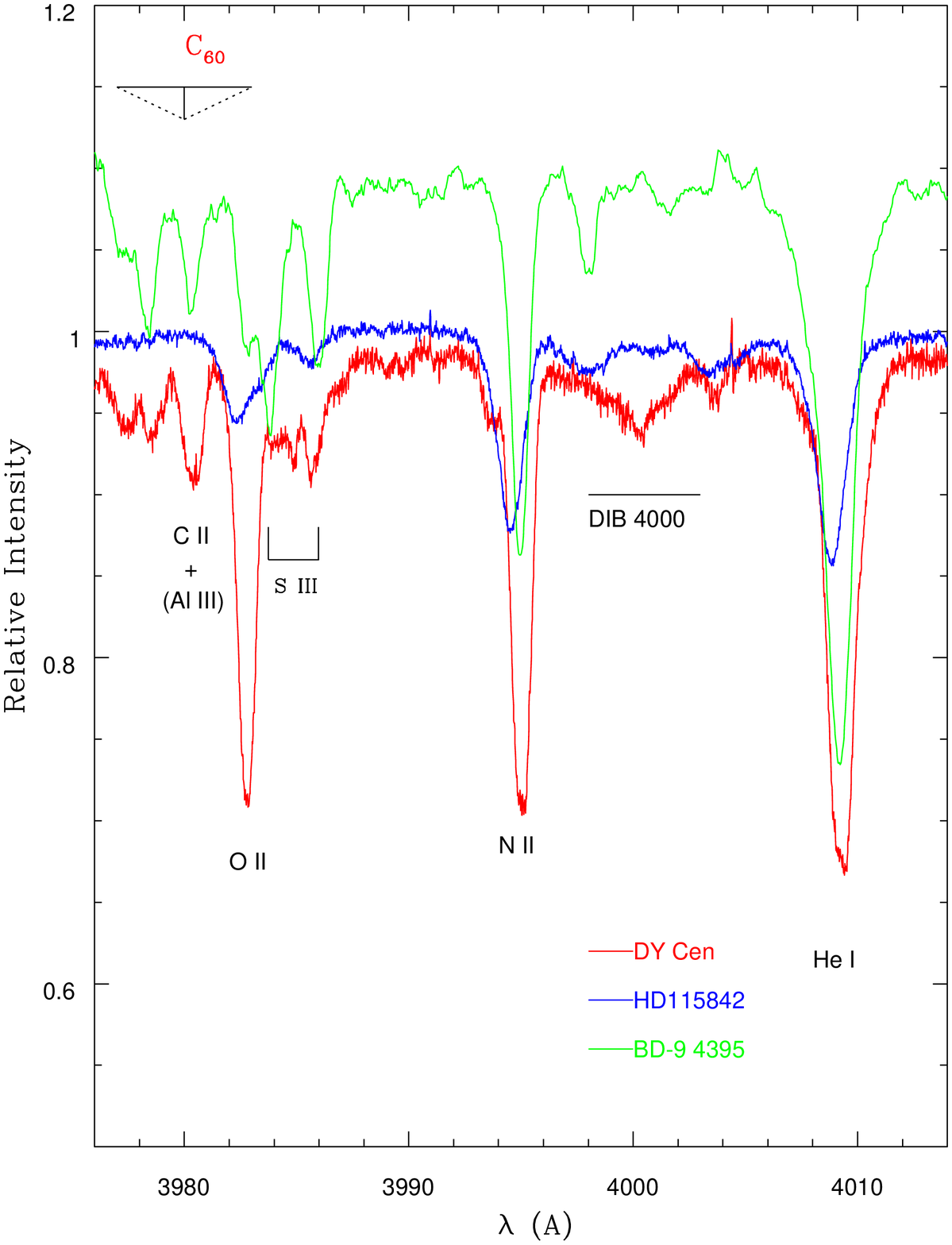}
\caption{The spectra of DY Cen (in red) and HD 115842 (in blue) around 3760 \AA\
(left panel) and 4000 \AA\ (right panel; where the eHe star BD $-9^\circ$ 4395
is also displayed in green) are shown. The expected positions of the C$_{60}$
features are marked on top of the spectra.. The FWHM of the features are
indicated by the dotted line. Note that there is no evidence (additional
absorption) in DY Cen for the presence of the neutral C$_{60}$ features at 3760,
3980, and 4024 \AA. However, there is an additional absorption band at 4000 \AA\
in DY Cen (marked by a dark line and written DIB4000), which is not present in
either HD 115842 or in  the  extreme He star BD $-9^\circ$ 4395 that is expected
to show similar  spectrum as DY Cen. \label{fig1}}
\end{figure}

\clearpage

\begin{figure}
\includegraphics[angle=0,scale=.65]{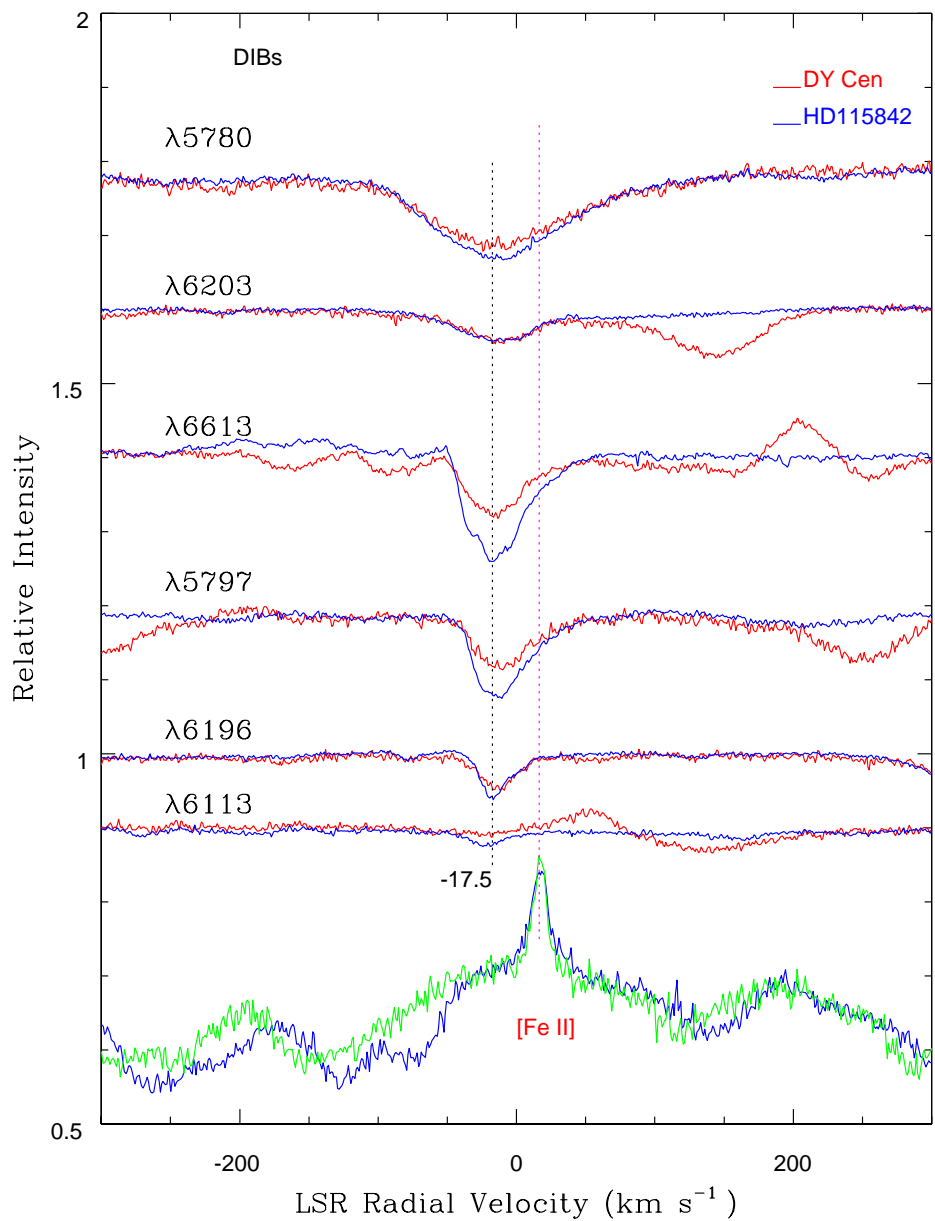}
\caption{Profiles of six DIBs of various strengths and widths are plotted  with
respect to the LSR radial velocity. Note that the profiles are shifted
vertically for clarity. The [Fe II] profile at the bottom shows the DY Cen's
systemic radial velocity. Note that the strengths and shapes of  5780 and  6203
\AA\ DIBs are almost identical for both stars. The other profiles show that DIBs
towards the DY Cen sight of line are weaker than towards HD 115842. To convert
the velocity scale to heliocentric, 4.6 and 5.1 km s$^{-1}$ should be added to
the LSR velocities of DY Cen and HD 115842, respectively. \label{fig3}}
\end{figure}

\clearpage

\begin{figure}
\includegraphics[angle=0,scale=.45]{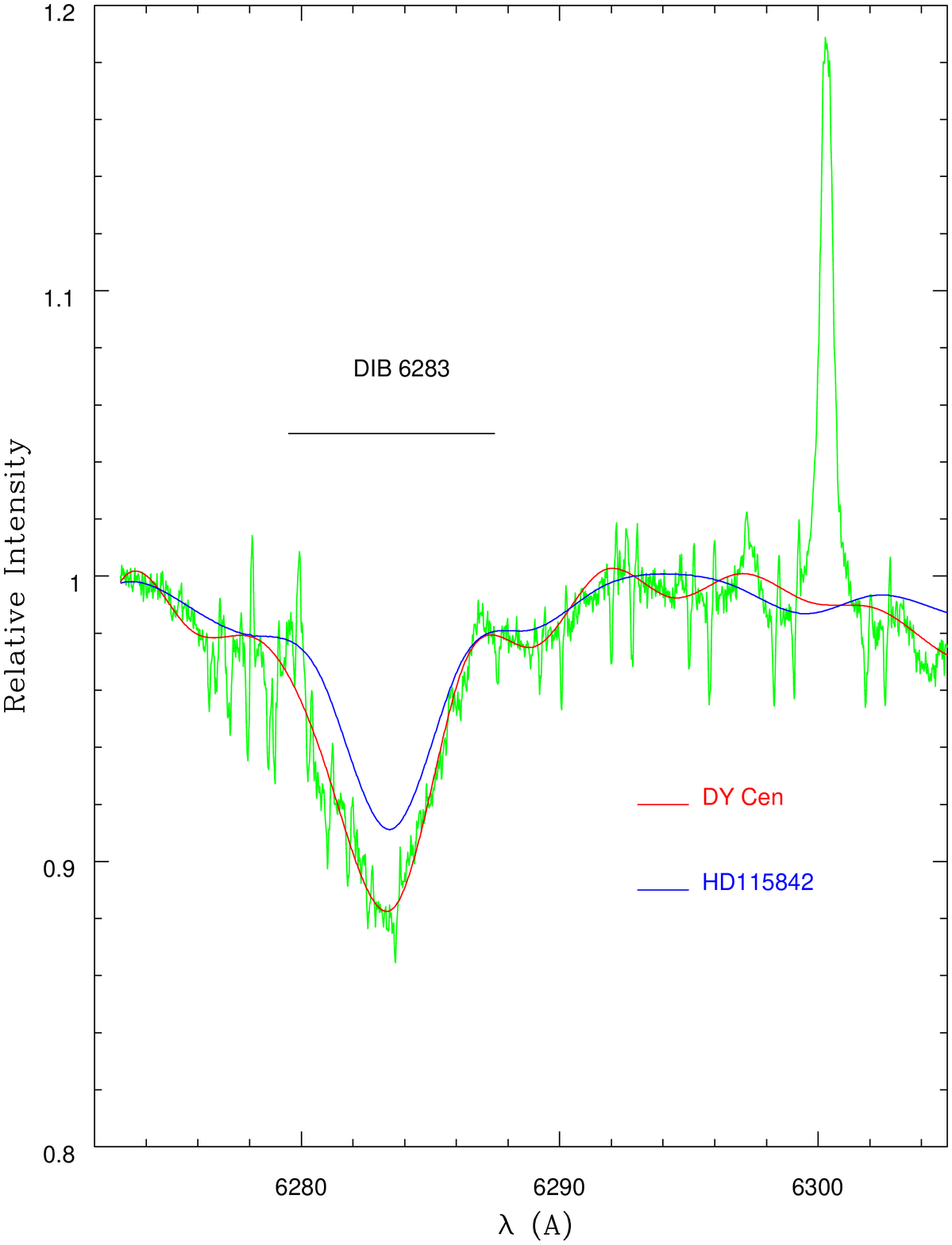}%
\includegraphics[angle=0,scale=.45]{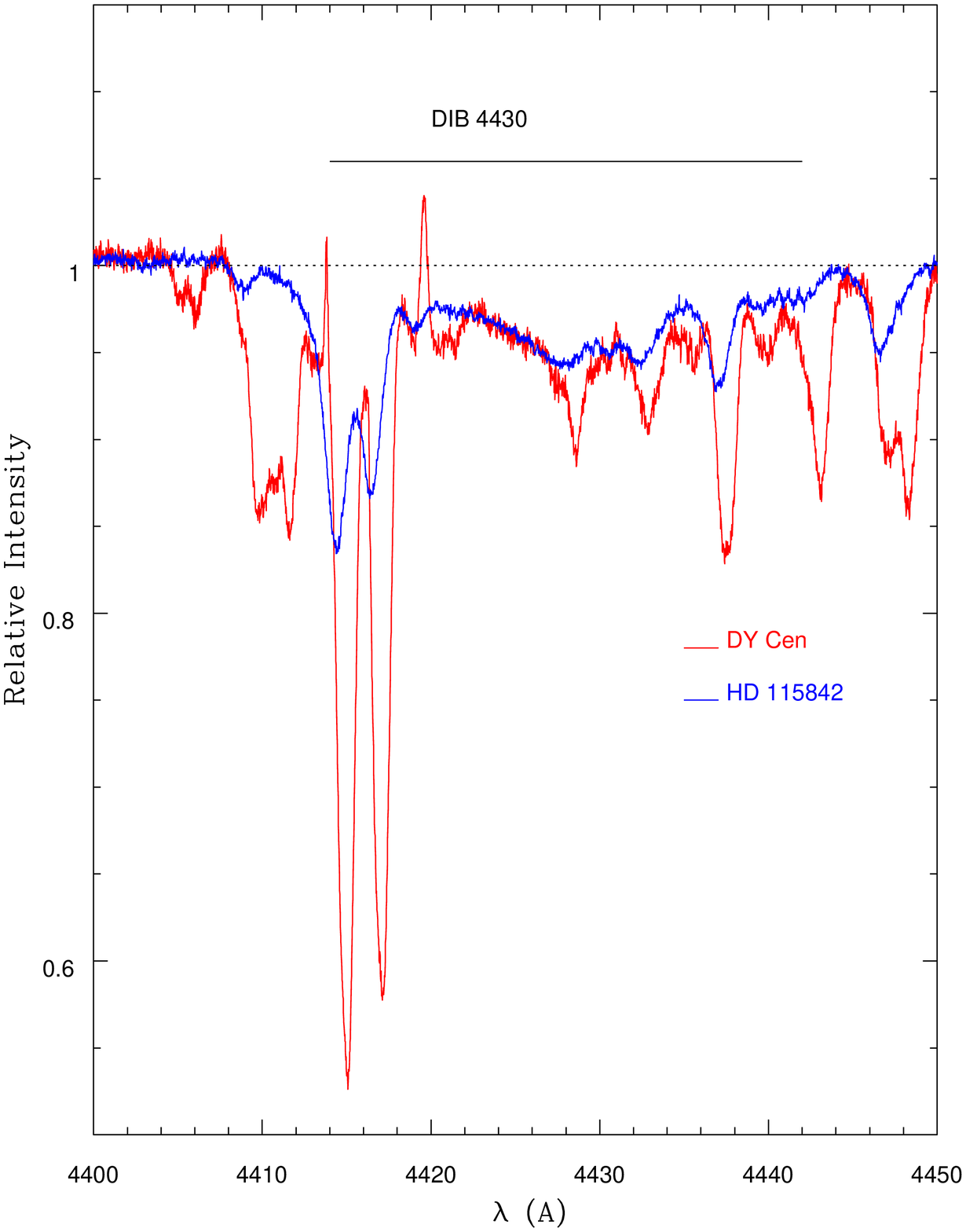}
\caption{Spectral regions around the DIBs at 6284 \AA\ (left panel) and
4428 \AA\ (right panel) in DY Cen (in red) and HD 115842 (in blue). The telluric
line corrected spectrum of both stars are shown by smooth lines (red and blue).
The not fully telluric corrected spectrum with emission of [O I] line in DY Cen
 is shown in green. It is obvious that
the 6284 \AA\ DIB is stronger in DY Cen spectrum than towards HD 115842
  while the DIB at 
4428 \AA\ (right panel) is of similar  strength in  both stars. \label{fig4}}
\end{figure}

\end{document}